\documentclass[superscriptaddress,altaffilletter,amsmath,amssymb,aps,twocolumn,floatfix]{revtex4-1}

\usepackage[english]{babel}
\usepackage[colorinlistoftodos]{todonotes}
\usepackage{multirow}
\usepackage{comment}
\usepackage{relsize}
\usepackage{float}
\usepackage{textcomp}
\usepackage{amsmath}
\usepackage[justification=justified]{subcaption}
\usepackage[justification=justified, figurename=FIG.]{caption}
\usepackage[compact]{titlesec}
\titlespacing{\section}{0pt}{3ex}{2ex}
\titlespacing{\subsection}{0pt}{2ex}{2ex}
\titlespacing{\subsubsection}{0pt}{1.5ex}{1.5ex}

\usepackage[hidelinks]{hyperref}

\DeclareCaptionLabelFormat{UC}{\MakeUppercase{#1} #2}
\usepackage{xcolor}
\usepackage{graphicx}
\usepackage{subcaption}
\usepackage{epstopdf}
\usepackage{color}
\usepackage{dcolumn}
\usepackage{bm}
\usepackage{hyperref}
\usepackage[mathlines]{lineno}
\usepackage{threeparttable}
\usepackage{booktabs} 
\usepackage{siunitx}
\usepackage{placeins}
\usepackage{xspace}

\begin{document}

\title{Sensitivity of nEXO to $^{136}$Xe Charged-Current Interactions: Background-free Searches for Solar Neutrinos and Fermionic Dark Matter}

\date{\today}
\newcommand{\UCSD}{\affiliation{Physics Department, University of California San Diego, La Jolla, CA 92093, USA}}
\newcommand{\McGill}{\affiliation{Physics Department, McGill University, Montr\'eal, QC H3A 2T8, Canada}}
\newcommand{\Stanford}{\affiliation{Physics Department, Stanford University, Stanford, CA 94305, USA}}
\newcommand{\Erlangen}{\affiliation{Erlangen Centre for Astroparticle Physics (ECAP), Friedrich-Alexander University Erlangen-N{\"u}rnberg, Erlangen 91058, Germany}}
\newcommand{\PNNL}{\affiliation{Pacific Northwest National Laboratory, Richland, WA 99352, USA}}
\newcommand{\Carleton}{\affiliation{Department of Physics, Carleton University, Ottawa, ON K1S 5B6, Canada}}
\newcommand{\UMass}{\affiliation{Amherst Center for Fundamental Interactions and Physics Department, University of Massachusetts, Amherst, MA 01003, USA}}
\newcommand{\ITEP}{\affiliation{National Research Center ``Kurchatov Institute'', Moscow, 123182, Russia}}
\newcommand{\LLNL}{\affiliation{Lawrence Livermore National Laboratory, Livermore, CA 94550, USA}}
\newcommand{\UK}{\affiliation{Department of Physics and Astronomy, University of Kentucky, Lexington, KY 40506, USA}}
\newcommand{\BNL}{\affiliation{Brookhaven National Laboratory, Upton, NY 11973, USA}}
\newcommand{\SLAC}{\affiliation{SLAC National Accelerator Laboratory, Menlo Park, CA 94025, USA}}
\newcommand{\RPI}{\affiliation{Department of Physics, Applied Physics, and Astronomy, Rensselaer Polytechnic Institute, Troy, NY 12180, USA}}
\newcommand{\Laurentian}{\affiliation{School of Natural Sciences, Laurentian University, Sudbury, ON P3E 2C6, Canada}}
\newcommand{\IHEP}{\affiliation{Institute of High Energy Physics, Chinese Academy of Sciences, Beijing, 100049, China}}
\newcommand{\IME}{\affiliation{Institute of Microelectronics, Chinese Academy of Sciences, Beijing, 100029, China}}
\newcommand{\Sherbrooke}{\affiliation{Universit\'e de Sherbrooke, Sherbrooke, QC J1K 2R1, Canada}}
\newcommand{\Alabama}{\affiliation{Department of Physics and Astronomy, University of Alabama, Tuscaloosa, AL 35405, USA}}
\newcommand{\UNCW}{\affiliation{Department of Physics and Physical Oceanography, University of North Carolina Wilmington, Wilmington, NC 28403, USA}}
\newcommand{\UBC}{\affiliation{Department of Physics and Astronomy, University of British Columbia, Vancouver, BC V6T 1Z1, Canada}}
\newcommand{\TRIUMF}{\affiliation{TRIUMF, Vancouver, BC V6T 2A3, Canada}}
\newcommand{\Drexel}{\affiliation{Department of Physics, Drexel University, Philadelphia, PA 19104, USA}}
\newcommand{\ORNL}{\affiliation{Oak Ridge National Laboratory, Oak Ridge, TN 37831, USA}}
\newcommand{\CSU}{\affiliation{Physics Department, Colorado State University, Fort Collins, CO 80523, USA}}
\newcommand{\Yale}{\affiliation{Wright Laboratory, Department of Physics, Yale University, New Haven, CT 06511, USA}}
\newcommand{\USD}{\affiliation{Department of Physics, University of South Dakota, Vermillion, SD 57069, USA}}
\newcommand{\Mines}{\affiliation{Department of Physics, Colorado School of Mines, Golden, CO 80401, USA}}
\newcommand{\CUP}{\affiliation{IBS Center for Underground Physics, Daejeon, 34126, South Korea}}
\newcommand{\UWC}{\affiliation{Department of Physics and Astronomy, University of the Western Cape, P/B X17 Bellville 7535, South Africa}}
\newcommand{\SUBATECH}{\affiliation{SUBATECH, Nantes Universit\'e, IMT Atlantique, CNRS/IN2P3, Nantes 44307, France}}
\newcommand{\Caltech}{\affiliation{California Institute of Technology, Pasadena, CA 91125, USA}}
\newcommand{\Bern}{\affiliation{LHEP, Albert Einstein Center, University of Bern, 3012 Bern, Switzerland}}
\newcommand{\Skyline}{\affiliation{Skyline College, San Bruno, CA 94066, USA}}
\newcommand{\SNOLAB}{\affiliation{SNOLAB, Lively, ON P3Y 1N2, Canada}}
\newcommand{\Muenster}{\affiliation{Institut f{\"u}r Kernphysik, Westf{\"a}lische Wilhelms-Universit{\"a}t M{\"u}nster, M{\"u}nster 48149, Germany}}
\newcommand{\FRIB}{\affiliation{Facility for Rare Isotope Beams, Michigan State University, East Lansing, MI 48824, USA}}
\newcommand{\Queens}{\affiliation{Department of Physics, Queen's University, Kingston, ON K7L 3N6, Canada}}
\newcommand{\Windsor}{\affiliation{Department of Physics, University of Windsor, Windsor, ON N9B 3P4, Canada}}
\newcommand{\TUM}{\affiliation{Physikdepartment and Excellence Cluster Universe, Technische Universit{\"a}t M{\"u}nchen, Garching 80805, Germany}}
\newcommand{\Montclaire}{\affiliation{Department of Physics and Astronomy, Montclair State University, Montclair, NJ 07043, USA}}
\newcommand{\McMaster}{\affiliation{Department of Physics and Astronomy, McMaster University, Hamilton, ON L8S 4M1, Canada}}
\newcommand{\SJTU}{\affiliation{School of Physics and Astronomy, Shanghai Jiao Tong University, Shanghai 200240, China}}
\newcommand{\Hawaii}{\affiliation{Department of Physics and Astronomy, University of Hawaii at Manoa, Honolulu, HI 96822, USA}}
\newcommand{\KEK}{\affiliation{KEK, High Energy Accelerator Research Organization 1-1 Oho, Tsukuba, Ibaraki 305-0801, Japan}}
\newcommand{\FBK}{\affilition{Fondazione Bruno Kessler, 38123 Trento, Italy}}

\author{G.~Richardson} \SLAC \Yale
\author{B.~G.~Lenardo}\SLAC
\author{D.~Gallacher}\McGill
\author{R.~Saldanha}\PNNL

\author{P.~Acharya}\Alabama
\author{S.~Al Kharusi}\Stanford
\author{A.~Amy}\SUBATECH\UMass
\author{E.~Angelico}\Stanford
\author{A.~Anker}\SLAC
\author{I.~J.~Arnquist}\PNNL
% ...
\begin{abstract}
\noindent

We study the sensitivity of nEXO to solar neutrino charged-current interactions, $\nu_e + ^{136}$Xe$\rightarrow ^{136}$Cs$^* + e^-$, as well as analogous interactions predicted by models of fermionic dark matter. Due to the recently observed low-lying isomeric states of $^{136}$Cs, these interactions will create a time-delayed coincident signal observable in the scintillation channel. Here we develop a detailed Monte Carlo of scintillation emission, propagation, and detection in the nEXO detector to model these signals under different assumptions about the timing resolution of the photosensor readout. We show this correlated signal can be used to achieve background discrimination on the order of $10^{-9}$, enabling nEXO to make background-free measurements of solar neutrinos above the reaction threshold of 0.668~MeV. We project that nEXO could measure the flux of CNO solar neutrinos with a statistical uncertainty of 25\%, thus contributing a novel and competitive measurement towards addressing the solar metallicity problem. Additionally, nEXO could measure the mean energy of the $^7$Be neutrinos with a precision of $\sigma \leq 1.5$~keV and could determine the survival probability of $^{7}$Be and \textit{pep} solar $\nu_e$ with precision comparable to state-of-the-art. These quantities are sensitive to the Sun’s core temperature and to non-standard neutrino interactions, respectively. Furthermore, the strong background suppression would allow nEXO to search for for charged-current interactions of fermionic dark matter in the mass range $m_\chi$~=~$0.668$--$7$~MeV with a sensitivity up to three orders of magnitude better than current limits.

\end{abstract}

\maketitle

\section{\label{sec:intro}INTRODUCTION}

Over the past several decades, liquid xenon (LXe) time projection chambers (TPCs) have achieved world-leading sensitivities in areas such as direct dark matter detection and searches for neutrinoless double-beta decay ($0 \nu \beta \beta$) \cite{LZ_first_results, XENON_results, PandaX_results, EXO-200}. Some of the benefits of LXe TPCs include their self-shielding and monolithic structure, as well as their scalability to achieve large exposures. Recent observations of low-lying isomeric states in $^{136}$Cs indicate that LXe TPCs, particularly enriched Xe detectors such as the proposed nEXO detector~\cite{NEXO_pCDR_2018}, also have the potential to expand their physics program through studies of $^{136}$Xe charged-current ($^{136}$Xe-CC) interactions of the form: 
\begin{equation}
\label{eq:CC}
    \nu_e + ^{136}\mathrm{Xe} \rightarrow ^{136}\mathrm{Cs}^* + e^-
\end{equation}

In this interaction, an incoming electron neutrino is absorbed by a $^{136}$Xe nucleus, depositing its energy into the detector in the form of an excited $^{136}$Cs nucleus preferentially in its lowest lying $J^{\pi}=1^+$ state with energy $E_{1^+}$= $589$~keV~\cite{Isomeric_Obs} and an electron. The electron will have energy $E_e=E_{\nu} - E_{thresh}$, where $E_{thresh} = E_{1^+} + Q=668$~keV is the reaction threshold in terms of total energy and $Q=79$~keV is the reaction $Q$-value~\cite{Ge:2023mnb}. 

\begin{figure}[t]
    \centering
    \includegraphics[width=1\columnwidth]{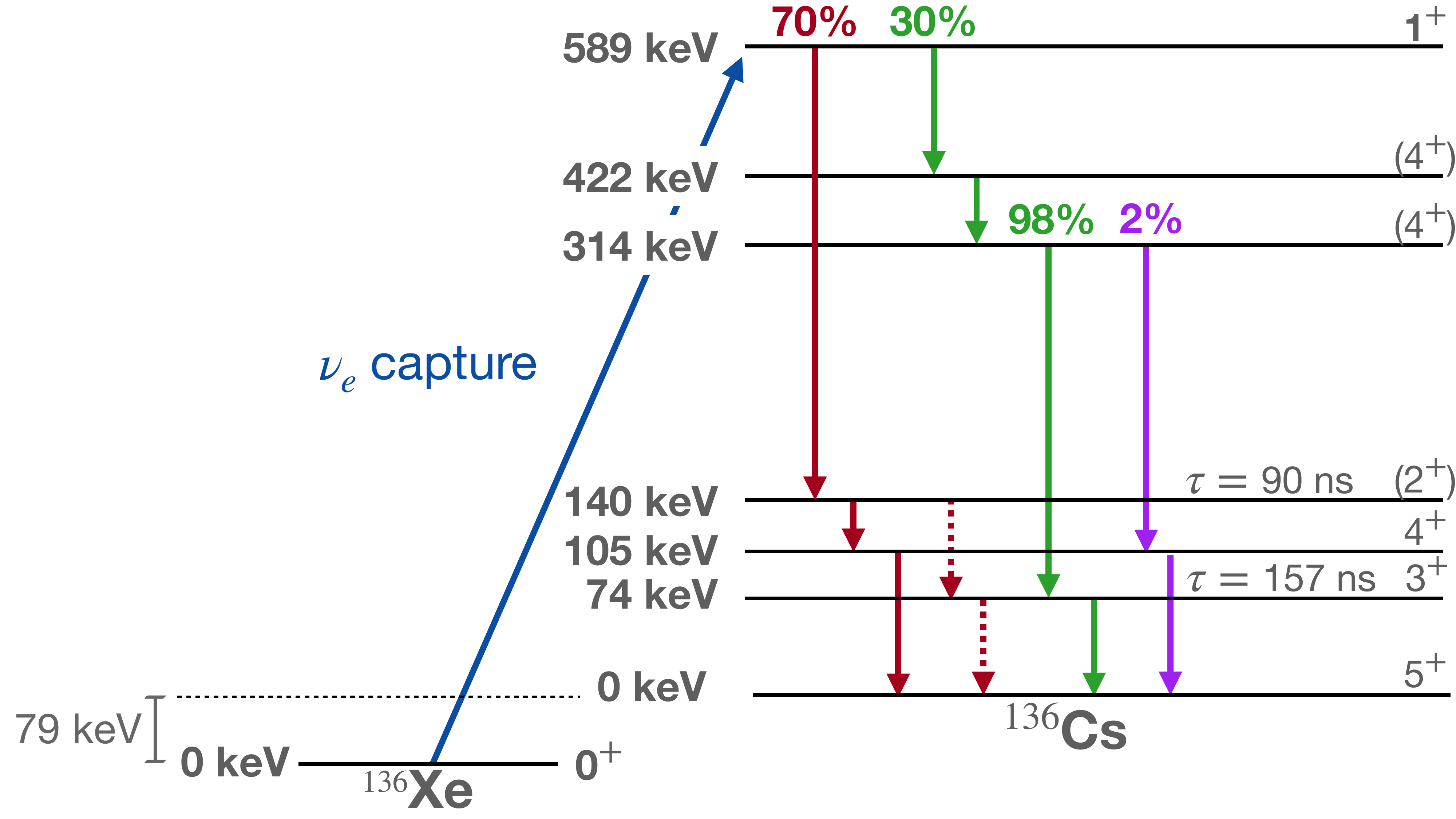}
    \caption{A simplified level diagram indicating the $^{136}$Xe-CC interaction and subsequent de-excitation of the $^{136}$Cs. $70\%$ of the decays from the $1^+$ state are expected to pass through the $140$~keV isomeric state (red) with some unknown fraction further passing through the $74$~keV isomeric state (red-dashed) creating a triple coincidence. The remaining $30\%$ are expected to de-excite via the $589-422-314$~keV transitions, with $98\%$ of those states passing through the $74$~keV isomeric state (green), and $2\%$ proceeding promptly to the ground state via the $105$~keV state (purple)~\cite{Isomeric_Obs, Wimmer}. Spin-parity assignments shown in parenthesis are currently under tension~\cite{Isomeric_Obs, Rebeiro}.}
    \label{fig:level_diagram}
\end{figure}

Following the interaction, the $^{136}$Cs nucleus will de-excite to either the $140$~keV isomeric state ($\sim 70$\% branching) with lifetime $90 \pm 5$~ns, or the $74$~keV isomeric state ($\sim 30$\% branching) with lifetime $157 \pm 4$~ns before finally transitioning to the ground state ($0.6\%$ of the $1^+$ states are expected to decay to the ground state through prompt transitions)~\cite{Isomeric_Obs}. These transitions can occur through either $\gamma$-ray de-excitation or internal conversion. From the ground state, the nucleus will eventually $\beta$-decay to $^{136}$Ba, however the half-life of this decay is sufficiently long ($T_{1/2} \approx$ 13 days) that it is not expected to be part of the same event~\cite{Cs_Beta_Decay_lifetime}. A partial level diagram is shown in Figure~\ref{fig:level_diagram}. The decay will deposit a total energy $E_{vis} = E_{\nu} - Q=E_e+E_{1^+}$ in the detector with a minimum of $E_{min}=E_{1^+}=0.589$~MeV.
The transition through an isomeric state will create a \textit{time-delayed coincident signal} which can be used to tag the interaction described by Eq.~\ref{eq:CC} and reject backgrounds. 

Double-beta decay isotopes, such as $^{136}$Xe, are ideal targets for this interaction; they are stable against single beta decay and the odd-odd daughter generally provides a complex low-energy level structure that allows for low interaction thresholds and the possibility of long-lived isomers~\cite{Raghavan:1997ad}. LXe $0 \nu \beta \beta$ detectors are well suited for this analysis for several reasons. These detectors are typically optimized for $\mathcal{O}$(MeV) energy deposits given $Q_{\beta \beta}^{136\mathrm{Xe}} = 2.45$ MeV. Since Eq.~\ref{eq:CC} is an absorption interaction, it will deposit the total neutrino energy, minus a small offset, into the detector, resulting in $\mathcal{O}$(MeV) energy deposits given the relatively high flux of neutrinos in this range (solar and supernovae)~\cite{neutrino_sources}. Furthermore, $0 \nu \beta \beta$ LXe TPCs often use xenon that has been isotopically enriched in $^{136}$Xe ($\sim 90\%$ $^{136}$Xe)~\cite{nEXO_sens}, providing significantly greater exposure per unit target mass to $^{136}$Xe-CC interactions than LXe TPCs optimized to search for dark matter which preferentially use natural xenon ($\sim$9\% $^{136}$Xe)~\cite{EXO-200, LZ_0nuBB_sensitivity2020}.

Previous phenomenological studies predicted the existence of isomeric states of $^{136}$Cs, suggesting that their characteristic time delays could enable background suppression factors as large as $\mathcal{O}(10^{10})$ in LXe TPCs~\cite{Haselschwardt:2020ffr}. Only recently have the relevant isomeric states been experimentally observed and their lifetimes measured~\cite{Isomeric_Obs}, making it possible to quantitatively assess the impact of this signature on experimental sensitivity. While prior experimental work has utilized the $^{136}$Xe-CC channel, none have been positioned to take advantage of the delayed coincidence signature~\cite{Ako, Sam}. In this study, we incorporate the newly measured isomer lifetimes into detailed Monte Carlo (MC) simulations using a realistic model of the nEXO detector to accurately project its sensitivity to these signals.

In particular, this work will study how the planned nEXO experiment can use $^{136}$Xe-CC ineractions to serve as a solar neutrino detector, combining the low background rate of radiochemical detection from isotopic specificity \cite{DAVIS199413, SAGE, Gallex} with the real-time measurement capabilities of liquid-scintillator/Cherenkov-based experiments \cite{Borexino_design, SNO, Theia}. Furthermore, this charged-current channel allows LXe TPCs like nEXO to directly measure the incoming neutrino energy rather than a broad recoil spectrum \cite{Gabriel, Haselschwardt:2020ffr}. This unique combination of abilities could allow nEXO to make measurements that complement existing and planned solar neutrino experiments and help address several important questions in the field of particle astrophysics~\cite{Borexino_design, Theia, JUNO, SNO+}.

One such open question in solar physics is the heavy element ($Z\geq 3$) composition of the Sun. This ``Solar Metallicity'' question arises from a discrepancy between helioseismological and spectroscopic solar measurements which has given rise to two contrasting sets of solar models: the high metallicity models (HZ) and the low metallicity models (LZ)~\cite{Vinyoles:2016djt}. A third method of measuring this composition, which might help resolve the discrepancy, is to directly measure the flux of neutrinos produced by the Carbon-Nitrogen-Oxygen (CNO) cycle within the Sun~\cite{Gabriel, Christensen_Dalsgaard_2021, Cerdeno:2017xxl}.

Another important question is the temperature of the Sun's core, which is primarily constrained using the $^8$B neutrino flux~\cite{PhysRevD.53.4202}. An independent measurement can be obtained by measuring the mean energy of the neutrinos produced from $^7$Be electron capture in the Sun's core. Unlike laboratory $^7$Be sources, which decay by electron capture from discrete atomic shells, solar $^7$Be captures electrons from a hot plasma continuum due to thermal ionization within the Sun's core. This is expected to induce a distortion and shift of approximately $1.3$~keV to the resulting neutrino spectrum compared to its terrestrial counterpart. By precisely measuring this line-shift, one can make a direct measurement of the average temperature of the solar core and check consistency with $^{8}$B-derived values \cite{Bahcall:1994cf}.

A final question where LXe TPCs could contribute through measurement of solar neutrinos is the existence of non-standard neutrino interactions (NSI) occurring in the Sun which could act as modifications on the Mikheyev–Smirnov–Wolfenstein Large Mixing Angle (MSW-LMA) description for neutrino oscillations in matter. By measuring the survival probability, $P_{ee}$, of electron solar neutrinos at Earth such effects can be either measured or constrained~\cite{Borexino_MSW, SNO, Gabriel}.

In addition to solar neutrinos, this work will also discuss how this detection channel will enable nEXO to search for fermionic dark matter (FDM) coupling to electrons through the addition of a new $\chi-W'-e^-$ vertex to the Standard Model (SM), where $\chi$ is the FDM particle and $W'$ a new dark sector boson. Because of this coupling, $\chi$ is capable of undergoing charged current interactions with SM particles, similar to that of Eq.~\ref{eq:CC}, though it may not preferentially excite the lowest lying $1^+$ state as neutrinos do~\cite{Dror:2019onn, Second_Dror}.

We begin with a brief overview of LXe TPCs and the proposed nEXO detector in Sec.~\ref{sec:nEXO}. In Sec.~\ref{sec:CC-Signal} the MC simulation developed for this work and the analysis performed on these MCs are discussed, giving estimates for the background and signal efficiencies of the $^{136}$Xe-CC channel. This is followed by a brief overview of the backgrounds expected in nEXO in Sec~\ref{sec:backgrounds}. In Sec.~\ref{sec:solar-nu} these results are applied to the case of solar neutrinos. Then, in Sec.~\ref{sec:DM}, projections are given for regions of FDM parameter space that the nEXO detector could probe. Finally, Sec.~\ref{sec:ktonne} discusses how these results could scale to future generations of LXe TPCs. 

\section{\label{sec:nEXO}THE NEXO DETECTOR}

The proposed nEXO detector, which we analyze in detail here, plans to deploy $5$ tonnes ($3.3$ fiducial tonnes) of xenon enriched to $90\%$ in $^{136}$Xe in a $1.3$~m square-cylindrical liquid xenon TPC. The experiment is designed to search for $0 \nu \beta \beta$ with a half-life sensitivity of $T^{0 \nu}_{1/2}>10^{28}$ years~\cite{nEXO_sens}. Particles interacting with LXe produce both scintillation and ionization signals which can be used to reconstruct the energy, position, topology and timing of events in the fiducial volume. The scintillation signal in nEXO, consisting of vacuum ultra-violet (VUV) photons, will be read out promptly by more than $7,000$ silicon photomultiplier (SiPM) channels placed around the barrel of the TPC. The ionization signal will be measured using approximately 4,000 charge readout channels tiling the anode plane at the top of the detectors \cite{zepeng_sim, Mike, nEXO_sens}.

The scintillation signal itself provides a measurement of the energy and time of the event. The timing profile of the scintillation signal, which is critical for the present work, is driven by three factors: 1) the scintillation time of LXe, which contains a fast ($\tau_s = 4$~ns) and a slow ($\tau_t = 26$~ns) component~\cite{Brian_LUX, XMASS, LXe_scint_1, LXe_scint_2}, 2) the travel time of photons in the detector which depends on the size of the detector, coverage of optical sensors, and reflectivity of surfaces, and 3) the timing resolution of the detector readout channels. 
nEXO has demonstrated multiple VUV-sensitive SiPMs meeting its requirements~\cite{PDE_req, AKO_SiPM, giacomo}; for this work the SiPMs are assumed to be consistent with the measurements of Hamamatsu Photonics HPK VUV4-Q-50 devices reported in~\cite{giacomo}. The details of nEXO's scintillation readout system are still under development, but timing resolution is expected to be in the range of $\sigma = 1$---$50$~ns. Early prototype systems have reported single-photon coincidence resolutions of $11$~ns~\cite{Brookhaven_SiPM}. In this work, we scan across different values of $\sigma$ to study its impact on the results, but choose an aggressive $\sigma = 1$~ns as our reference point. The electronics noise in the scintillation readout channel of nEXO is expected to be sufficiently low to achieve a signal-to-noise ratio for single photo-electrons (p.e.) of $>10/1$, firmly placing nEXO in the photon-counting regime. 

The ionization signal provides a complementary measurement of the event's energy, as well as the event's $X/Y$ position, topology, and, in conjunction with the scintillation signal, $Z$ position. The time profile of the ionization signal is determined by the longitudinal diffusion of the electron cloud and the drift velocity of the electrons in the LXe~\cite{EXO-diffusion, diffusion, Darwin_Drift}. 

By measuring the energy of events through both the scintillation and ionization channels, LXe TPCs are able to achieve energy resolution on the order of $\sigma / E \sim1$\% at the MeV scale~\cite{LZ_WIMP_Sensitivity2019, XENON_e_res, nEXO_sens}. The prompt energy of a $^{136}$Xe-CC signal is expected to be sufficiently higher than the threshold of the nEXO detector so that $^{136}$Xe-CC signals can be triggered on the prompt event, with the smaller delayed event being readout as part of the same event window~\cite{nEXO_sens}. 

nEXO's proposed location is SNOLAB which would provide a $2$~km overburden of rock, reducing the flux of cosmic-ray muons by more than a factor of $10^7$ compared to the surface. In order to further mitigate radioactive backgrounds, nEXO will employ a nested detector scheme. The center of nEXO will be the LXe TPC, which will be surrounded by a spherical cryostat containing $33$~tonnes of HFE-7200 which will act as both thermal insulation and neutron shield. 
The outermost layer will be a cylindrical volume of height $12.8$~m and diameter $12.3$~m containing ultrapure water that will act as both a radiation shield and an active muon veto system. The majority of muons which are expected to produce events within the LXe are those that pass within $3.35$~m of the TPC due to the transverse scale of the hadronic showers produced. At these distances, simulations predict that the nEXO muon veto system will be $>95$\% efficient~\cite{Soud_thesis}.

\section{{\label{sec:CC-Signal}}$^{136}$Xe-CC SIMULATION AND ANALYSIS}

\begin{figure*}[t]
    \centering
    \includegraphics[width=1\linewidth]{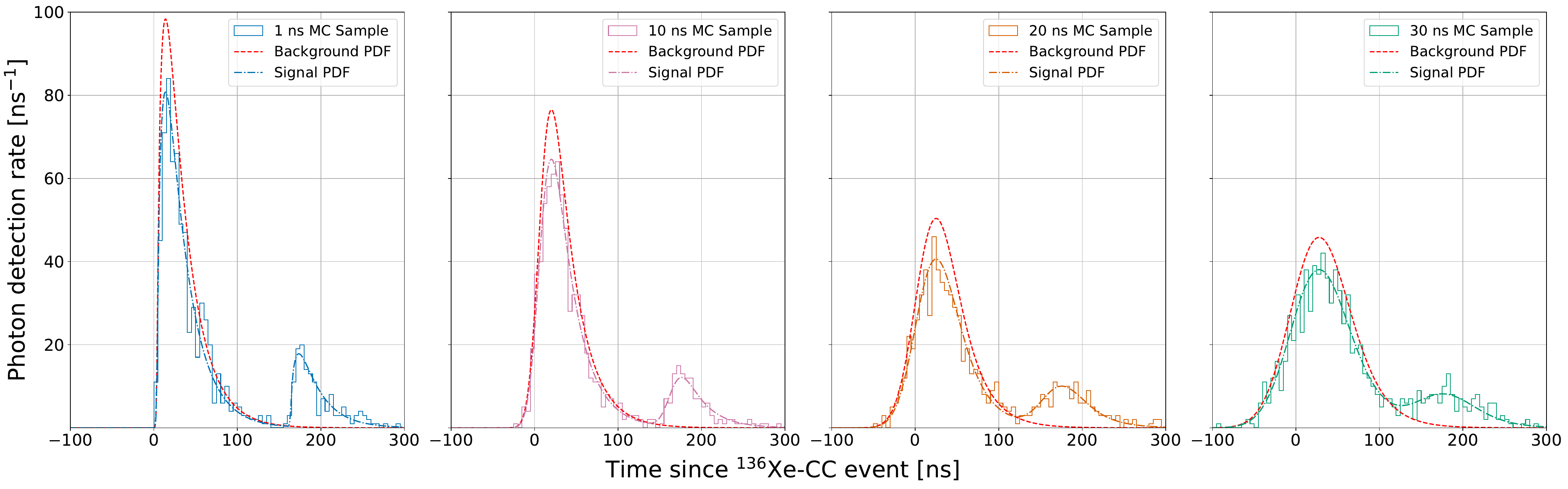}
    \caption{Example scintillation MC samples for a $157$~ns decay time along the $74$~keV decay path with incoming energy $E_{\nu}=668$~keV and detector timing resolutions $\sigma=$ $1$, $10$, $20$, and $30$~ns. The best fit signal PDF is shown for each MC sample along with the best fit background PDF consisting of a single scintillation pulse with no delayed component.}
    \label{fig:MC-sample}
\end{figure*}

To estimate nEXO's ability to use the time-delayed coincident signal produced by $^{136}$Xe-CC interactions, we have developed MC models that enable detailed simulations of both the scintillation and ionization signals that would be detected by nEXO. As fast timing is the most important handle for tagging these interactions, our scintillation model includes realistic detector effects and timing resolution. 

\subsection{Scintillation}
\label{sec:scintillation}

The scintillation model begins with a probability distribution function (PDF) that models the time profile of xenon scintillation, which takes the form  
\begin{equation}
    \label{eq:scintillation}
    S(t) = \frac{p_{s}}{\tau_s}e^{\frac{-t}{\tau_s}} + \frac{p_t}{\tau_t}e^{\frac{-t}{\tau_t}}
\end{equation}
where $p_s=0.05$ and $p_t=(1-p_s)$ are the probabilities of electrons and $\gamma$-rays exciting the singlet or triplet Xe excimers respectively and $\tau_s = 4$~ns and $\tau_t = 26$~ns are the corresponding lifetimes of the excimers~\cite{Brian_LUX, XMASS, LXe_scint_1, LXe_scint_2}.
This scintillation emission PDF is then convolved with the photon travel time PDF obtained by simulating the propagation of $10^6$ photons for events uniformly sampled from nEXO's fiducial volume using the Chroma simulation package \cite{chroma, nEXO_sens, Ako_thesis} and fitting the simulated data with an empirical analytic model of the form  
\begin{equation}
\label{travel time}
    T(t;t_0=0) = A  \delta(t) + \frac{B}{\tau_B}e^{\frac{-t}{\tau_B}} + \frac{C}{\tau_C} e^{\frac{-t}{\tau_C}} + \frac{D}{\tau_D}e^{\frac{-t}{\tau_D}}
\end{equation}
for computational ease. The normalization factor for this PDF has been absorbed into the amplitude parameters of the fit.
Finally, the detector timing resolution, $\sigma$, was included by convolving the resulting photon distribution with a normalized Gaussian function, $G$($t\mathrm{;} \sigma$), to create a detector response function, $R(t; \sigma)$.
\begin{equation}
\label{travel time}
    R(t; \sigma) = S(t) * T(t) * G(t; \sigma)
\end{equation}
To study the effect of detector timing resolution on the results, resolutions from $\sigma = 1$~ns to $\sigma = 50$~ns were considered. 

With Eq.~\ref{travel time} to model the shape of a single scintillation pulse, the PDF of the detector scintillation signal for a $^{136}$Xe-CC event, $P(t)$, is then created by adding together two response functions to model both the primary interaction and the delayed transition from the isomeric state:
\begin{equation}
\label{eq:PDF}
    P(t) = \frac{1}{N_p+N_d} \cdot [N_p\cdot R(t - t_0) + N_d \cdot R(t - t_0 - t_d)] 
\end{equation}
where the number of scintillation photons produced in the prompt and delayed pulse are given by $N_p$ and $N_d$, the time of the event by $t_0$, and the time of the delayed $^{136}$Cs de-excitation by $t_d$. $N_p$ and $N_d$ are modeled by randomly sampling a decay path of $^{136}$Cs using the branching ratios given in Ref.~\cite{Isomeric_Obs}, then using the corresponding prompt/delayed energies to simulate the numbers of scintillation photons with the Noble Element Simulation Technique~\cite{NEST}, which models scintillation yields and fluctuations. The average photon detection efficiency (PDE) of nEXO's fiducial volume, given in Ref.~\cite{nEXO_sens}, is then used to compute the number of detected photons in each event.
Due to limitations in the nuclear data, our simulations only consider cases where a single delayed signal comes from the $140$~keV state \textit{or} the $74$~keV state. This conservatively omits the possibility of transitions between the two isomeric states, which would create two delayed signals and would be more readily separated from backgrounds~\cite{Isomeric_Obs}.

\begin{figure*}[t]
    \centering
    \includegraphics[width=1\linewidth]{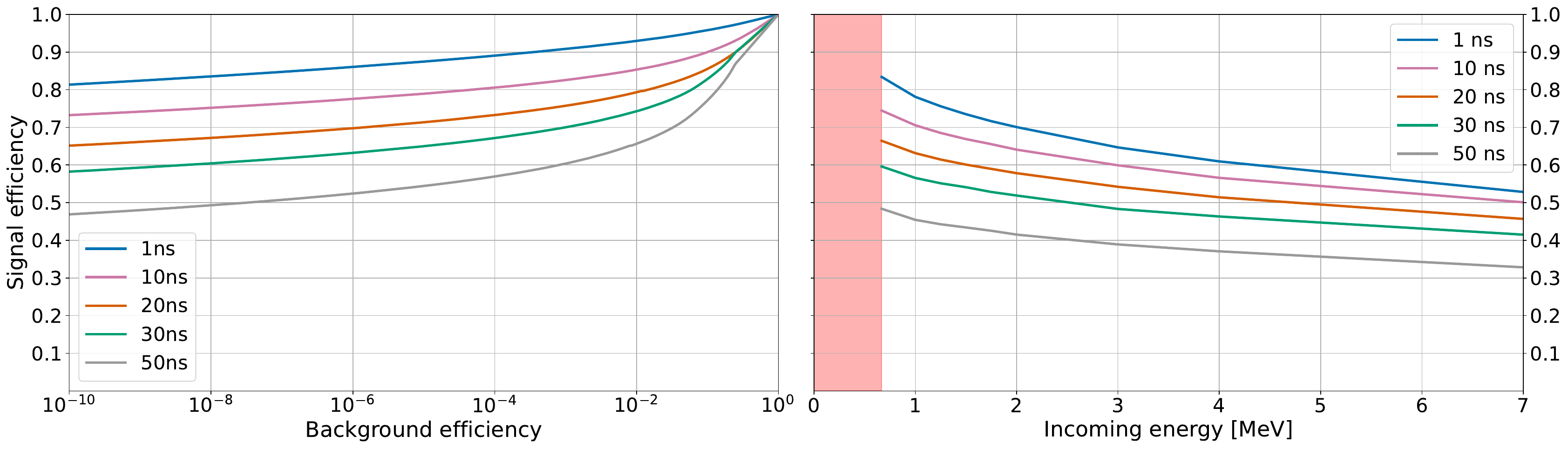}
    \caption{(left) The calculated ROC curve for the $\mathrm{^{136}Xe}$-CC detection channel in nEXO with incoming energy $E_{\nu} = 668$~keV. (right) The expected signal efficiency in nEXO with the $\mathrm{^{136}Xe}$-CC detection channel at $7.8 \cdot 10^{-9}$ background efficiency as a function of incoming energy.}
    \label{fig:ROC_effic}
\end{figure*}

In addition to the primary scintillation photons discussed above, secondary p.e. signals can also be created through instrumental effects. UV-induced light emission from passive materials, i.e. the delayed fluorescence of detector components, is expected to be negligible; in nEXO's open-field-cage design~\cite{nEXO:2020mec}, the only non-metallic materials visible to the active volume are pure silicon, fused silica, and thin MgF$_2$ films used in reflective coatings~\cite{Liu:2022hti}, none of which exhibit significant fluorescence in their pure form. We therefore assume that the dominant sources of secondary photon signals in nEXO are SiPM processes such as correlated avalanches (CAs), in which an avalanche from an absorbed photon creates a second avalanche in the same device, and external cross-talk (ExCT), in which a primary avalanche results in the emission of a photon that exits the device and can be detected by another channel~\cite{giacomo}. 
The probability that a detected photon causes a delayed correlated avalanche, or ``after pulse'', has been measured to be $P_{AP} \approx 10\%$ in Ref.~\cite{giacomo} and was modeled in the MC with a binomial distribution. A timing PDF for these after-pulses was calculated from the data in Ref.~\cite{giacomo} using the method in Ref.~\cite{Butcher} and sampled in the MC. The charge contained in each after-pulse, $Q_{AP}(t)$, was determined using the recharge rate reported in Ref.~\cite{giacomo}. 
The average number of ExCT photons emitted into LXe has been measured to be $1.23\pm0.43$ photons per avalanche by Ref.~\cite{Raymond} and $0.5^{+0.3}_{-0.2}$ photons per avalanche by Ref.~\cite{david}. In this work, we assume a weighted average of these two measurements of $\lambda_{ExCT}$ = 1 photon per avalanche. Because ExCT photons are in general not VUV, a separate Chroma simulation was performed in which $10^6$ photons were propagated from the face of SiPMs spread across the barrel of nEXO with wavelengths and opening angles given by Ref.~\cite{david}. The average travel time distribution of ExCT photons in nEXO was extracted from this simulation, and the probability of an ExCT photon being detected in nEXO given an initial SiPM avalanche was found to be $p_{ExCT} \approx 3\% $. The production and detection of ExCT photons in nEXO was modeled in the MC using a Poisson and binomial process and the detection time was sampled from the simulated PDF.
To calculate the production of delayed ExCT (dExCT)---ExCT generated from an after-pulse---it was assumed that the average number of dExCT photons produced in the LXe scales linearly with the charge in the avalanche so that $\lambda_{dExCT}(t) = \lambda_{ExCT}\frac{Q_{AP}(t)}{\lim_{t'\to \infty}Q_{AP}( t')}$. After being produced in the MC, dExCT photons were propagated using the same method as ExCT photons.

Finally, a single p.e. cut is applied to each channel to ensure proper timing reconstruction of the event---this removes channels that detect more than one photon as well as those which contain after-pulses.

Using this MC, we evaluate nEXO's ability to identify charged-current interactions using delayed coincidence signals in the scintillation channel. We generate samples of both signal events (i.e. charged-current interactions with delayed coincidences) and background events (events in which there is no delayed transition) with energies from $668$--$7000$~keV. Example events can be seen in Figure~\ref{fig:MC-sample}, with varying values of the time resolution, $\sigma$. Our discriminator uses a likelihood fit with three parameters ($t_d$ the decay time of the $^{136}$Cs, $t_0$ the event time, and $R$ the ratio of $N_p$ to $N_d$) performed on each MC event, and we calculate the log likelihood ratio of the event where $t_0$ and $R$ are considered nuisance parameters and profiled out~\cite{Cowan}. The likelihood score then provides a discriminant on which we can cut to separate signal-like events from background-like events.

These results were used to create the receiver operating characteristic (ROC) curve shown in the left panel of Figure~\ref{fig:ROC_effic}, which illustrates how the efficiency for detecting $^{136}$Xe-CC signals changes as a function of the background rejection efficiency. The time-delayed coincident scintillation signal produced by $^{136}$Xe-CC interactions enables background discrimination on the order of $10^{-9}$ with corresponding signal efficiencies of approximately $75\%$. In nEXO, this is sufficient discrimination power to operate at the level of $< 0.5$ expected background events in 10 years of livetime~\cite{nEXO_sens}.

\subsection{Ionization}

Our simulation of the ionization signals uses the Geant4-based \texttt{nexo-offline} framework developed in Refs.~\cite{nEXO_sens, zepeng_sim}. This code produces waveform-level simulated signals which include charge transport (i.e. drift, diffusion, and induction/collection) as well as models of the shaping and noise of the readout electronics. 

Resolving the time structure of $^{136}$Xe-CC events using the ionization channel will not be possible due to the much slower $\sim$mm/$\mu$s-scale electron drift velocity in liquid xenon~\cite{EXO-diffusion}. However, LXe TPCs are designed to determine the position of energy depositions with resolution of $\sim5$~mm in the X-Y direction and up to $0.7$~mm in the Z direction \cite{XnT_sensitivity:2020, LZ_first_results}, allowing these detectors to reconstruct the event topology. This ability permits classification of events as multi-site (MS) (events with multiple reconstructed interaction points within the detector) and single-site (SS) (events with a single reconstructed interaction point).

The electron and delayed de-excitation signal produced in a $^{136}$Xe-CC interaction are expected to constitute a single interaction point. A subset ($<5$\%) of the prompt de-excitations of the $^{136}$Cs produced are expected to proceed through internal conversion, in which case the event will often be SS. However, the majority of $^{136}$Xe-CC events ($>95$\%) are expected to de-excite via $\gamma$-emission~\cite{bric}, with the resulting $\gamma$-ray traveling on average $\sim 3$~cm from the primary event location, creating a MS event. In contrast, the dominant background, the double-beta decay ($2 \nu \beta \beta$) of $^{136}$Xe, will tend to create SS events, providing a separate handle in the ionization channel with which to discriminate against backgrounds~\cite{nEXO_sens}.

This discrimination was verified explicitly by simulating $10^6$ $2 \nu \beta \beta$ and $^{136}$Xe-CC events within the fiducial volume of nEXO with energy roughly equal to the unweighted average solar neutrino energy expected in this analysis, $E_{vis} = 1.25$~MeV \cite{zepeng_sim}. The Geant4 nuclear de-excitation data for $^{136}$Cs was modified to reflect the recent level scheme proposed by Ref.~\cite{Isomeric_Obs}. 

Various clustering algorithms were applied to these simulated data sets to separate the MS signals from the SS backgrounds. Although these algorithms were able to separate signal and background, their discrimination power was insufficient to provide any benefit to the analysis over simply making more aggressive selection cuts on the scintillation channel alone. For this reason, the ionization channel in this analysis was only used for energy and position reconstruction, but was not used for timing or topological discrimination. 

\section{\label{sec:backgrounds}BACKGROUNDS AND SYSTEMATICS}

\subsection{Backgrounds}

The primary background for $^{136}$Xe-CC events are interactions with no delayed emission, but which have a statistical fluctuation in photon detection time causing them to pass the likelihood-based selection criteria. These backgrounds in nEXO change significantly depending on the energy range. A comprehensive background model in the range from $1$~MeV to $3.5$~MeV is reported in Ref.~\cite{nEXO_sens}. Between $0.6$--$2.5$~MeV, backgrounds are dominated by the $2\nu \beta \beta$ decay of $^{136}$Xe and $\gamma$-rays from the decay of trace $^{40}$K in the detector components. In the mid-energy range, $2.5$--$5$~MeV, the backgrounds are significantly lower and dominated by other radioactive components in the detector, such as de-excitation $\gamma$-rays from the $^{232}$Th decay chain. At higher energies, $5$--$7$~MeV, experience with the EXO-200 detector predicts a low-intensity, featureless background spectrum dominated by ($n$,$\gamma$) activation~\cite{EXO-200-cosmogenic}. This activation can originate from multiple sources including cosmogenic muons, spontaneous fission, and ($\alpha$, $n$) reactions originating from exposure to Rn and dust during construction and assembly. We assume installation procedures will be designed to keep the generation of radioactivity-induced neutrons at acceptable levels so that cosmogenic activation will be the dominant source of such backgrounds. Under this assumption, we expect $\mathcal{O}(1)$ event at these energies in the lifetime of nEXO.
In total, between $589$~keV--$7$~MeV, nEXO expects $(6.4 \pm 0.4) \times 10^{7}$ background events in $10$ years of runtime ($1.9 \pm 0.1 \times 10^6$~counts/(ton$\cdot$yr)), where the vast majority of these events are $2 \nu \beta \beta$ and $^{40}$K decays and occur between $589$~keV and $2.4$~MeV \cite{nEXO_sens}. 

A second class of background events are those which create a false delayed-coincidence signal either by pile-up (multiple unrelated background events occurring in proximity to one another in both time and space, mimicking the time-delayed coincident signal of $^{136}$Xe-CC interactions), or by the radioactive decay of nuclei with de-excitation patterns similar to that of $^{136}$Cs. 

To derive an upper limit on the pileup rate, the problem is made both simpler and more conservative by ignoring the energy of the prompt de-excitation signal and only addressing the rate at which events of comparable energy to the delayed signal occur close enough in time and space as to be mistaken for the $^{136}$Xe-CC signal. The primary source of such low-energy backgrounds in nEXO is expected to be $2\nu \beta \beta$ events, as well as the $\beta$-decay of $^{85}$Kr ($Q_{Kr} = 687$~keV). Based off EXO-200 measurements, we estimate a $^{85}$Kr contamination rate of $(1.8 \pm 0.2)\cdot 10^3$ atoms per kg of Xe~\cite{LZ_WIMP_Sensitivity2019, EXO-200_Xe134}. Taking a time window of $5 \langle \tau_{Cs} \rangle$ = $550$~ns, a spatial window of $5$~cm, and assuming $5\%$ energy resolution at the delayed energies, the expected number of coincident backgrounds in nEXO is $<10^{-5}$ events per 10 years ($< 3 \times 10^{-7}$~events/(ton$\cdot$yr) \cite{Haselschwardt:2020ffr, Isomeric_Obs, nEXO_sens}. Even under these conservative assumptions, the pile-up rate is subdominant to the backgrounds arising from fluctuations in the photon detection times.

In terms of background events with a real CC-like coincidence signature, the most common source of delayed signals expected in nEXO are: 1) the $\beta$-decay of $^{85}$Kr which has a $0.4\%$ branching ratio to an isomeric state of $^{85}$Rb with half-life $T_{1/2} = 1015$~ns and energy $518$~keV, and 2) the $\beta$-decay of $^{212}$Bi ($Q_{Bi}=2.25$~MeV) which will feed the ground-state of $^{212}$Po which itself will $\alpha$-decay to $^{208}$Pb with a Q-value of $8.9$~MeV and half-life $T_{1/2}=299$~ns. Both signals can be rejected by the energy of the delayed component, which is significantly higher than the $^{136}$Cs isomeric transitions.
In addition, background events with a delayed structure can also arise from cosmogenic activation where spallation and fission products from a muon incident on the detector create unstable isotopes which may decay through isomeric transitions. Such potential backgrounds must not only have similar de-excitation structure to $^{136}$Cs (energy and lifetime), but must also be produced in the LXe and be fed by another decay process with sufficiently long half-life to escape nEXO's veto system ($T_{1/2} > \sim25$~min) , but also sufficiently short as to not be removed by nEXO's purifier ($T_{1/2} < \sim2$~days). We investigate this using dedicated Geant4 simulations of cosmogenic production and spallation in nEXO~\cite{Soud_thesis}, and cross-reference the results with a list of isotopes with isomeric transitions from the NNDC database~\cite{ensdf:1992}. We restrict this search to isomers with energies between $37$--$210$~keV and half-lives between $1$--$1000$~ns. We estimate that the total rate of such decays in the LXe volume will be less than $0.05$~events in $10$~years of runtime, making this background subdominant to the statistical background. 

\begin{figure*}
    \centering
    \includegraphics[width=\textwidth]{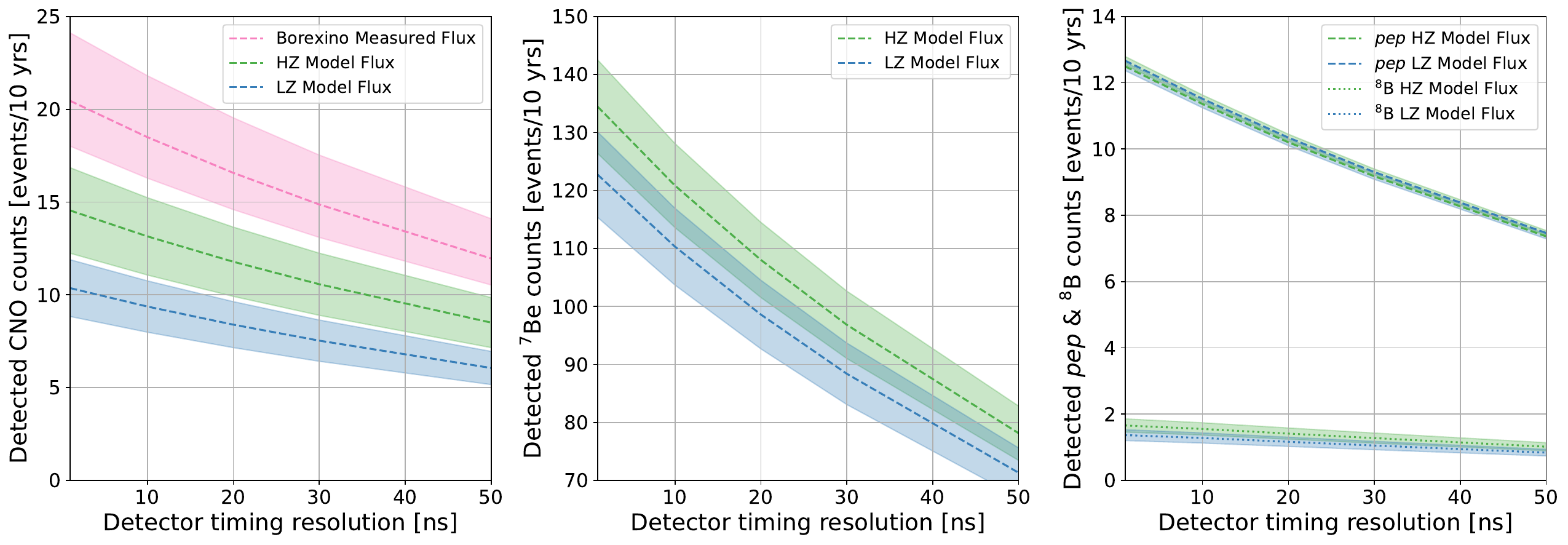}
    \caption{The expected number of solar neutrino events in nEXO for CNO (left) $^7$Be (middle) and $pep$ and $^8$B (right). Curves are shown for all species in the case of the HZ and LZ model~\cite{Gabriel}, and for Borexino's measured flux in the case of CNO~\cite{Borexino}. The colored bands indicate the uncertainty in the neutrino flux for each model or measurement. The Borexino measurements for $^7$Be, $pep$, and $^8$B fluxes agree well with HZ solar models and are omitted for clarity~\cite{BOREXINO_pp}.} 
    \label{fig:nEXO_solar_nu_rates}
\end{figure*}

\subsection{Systematic uncertainties}

The primary sources of systematic uncertainty expected in a $^{136}$Xe-CC analysis come from the uncertainty in overall solar neutrino flux on earth ($1-20\%$ relative uncertainty)~\cite{Gabriel} and the uncertainty in the Gamow-Teller transition strength ($\sim$11\%), which is measured by charge-exchange reaction experiments~\cite{GT, transitions} and is included in the CC cross section~\cite{Ejiri2019}. In this work, we assume that by the time the nEXO detector would be built and have collected 10 years of data, measurements of this transition strength could have advanced sufficiently to reduce the latter uncertainty to negligible levels~\cite{Puppe:2011zz}. Should this not be the case, the $^{136}$Xe-CC measurement of the $^7$Be neutrino flux could serve as an in-situ calculation of this transition strength and reduce the uncertainty to $\sim 9\%$ for the measurement of other components of the solar neutrino spectrum~\cite{Gabriel, BOREXINO_pp}.

\section{\label{sec:solar-nu} APPLICATION: SOLAR NEUTRINOS}

\begin{figure}[t]
    \centering
    \includegraphics[width=1\linewidth]{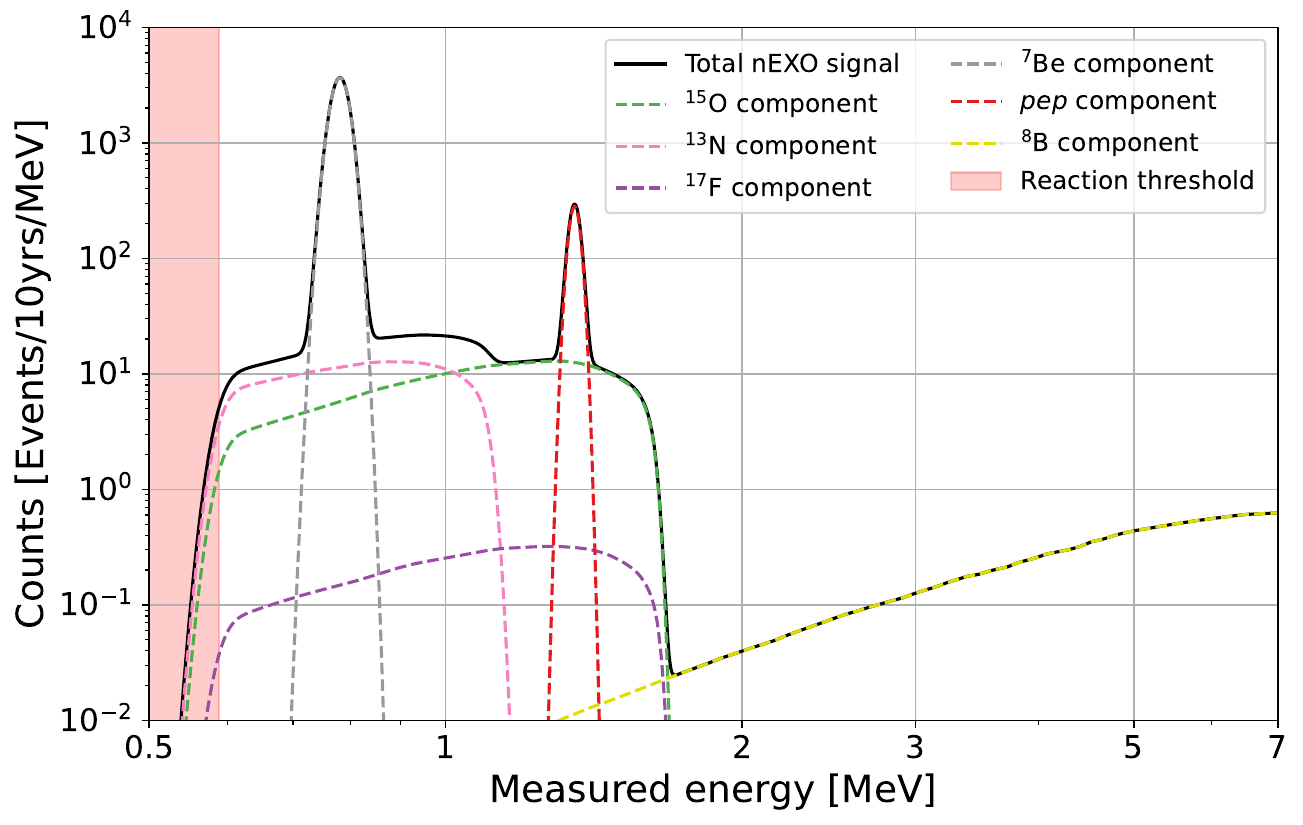}
    \caption{The expected solar neutrino spectrum in nEXO. The fluxes and spectra used in these simulations were taken from Ref.~\cite{Gabriel}.}
    \label{fig:solar-nu-spectra}
\end{figure}

To date, solar neutrino studies for LXe TPCs have focused on the detection of higher-energy $^8$B solar neutrinos through coherent elastic neutrino-nucleus scattering (CE$\nu$NS) \cite{XENON_cevens,panda_cevens} or of the higher-flux $pp$ solar neutrinos through electron scattering \cite{Panda_ER_pp, XLZD}. Here we show that the low reaction threshold of $^{136}$Xe-CC interactions allows this detection channel to measure low-energy components of the solar neutrino spectrum such as CNO, $^7$Be, and $pep$ neutrinos with background discrimination sufficient to reject the otherwise-overwhelming $2 \nu \beta \beta$ background which has limited the projected sensitivity in previous electron-scattering-based studies~\cite{CNO_depletion}. The expected solar neutrino spectrum in nEXO is shown in Figure~\ref{fig:solar-nu-spectra}.

The $^{136}$Xe-CC interaction rate over the energy range $E_i$ to $E_f$ is given below in units of interactions per second:

\begin{equation}
    \label{eq:flux}
    \Phi(E_i,E_f) = N_{\mathrm{Xe}} \int_{E_i}^{E_f}  \phi_{\nu}(E)  P_{ee}(E) \sigma(E) \epsilon(E)  dE 
\end{equation}
 where $N_{\mathrm{Xe}}$ is the number of $^{136}$Xe atoms in the active volume, $\phi_{\nu}(E)$ the total solar neutrino flux on Earth~\cite{Gabriel, Borexino}, $P_{ee}(E)$ the electron neutrino survival probability at Earth~\cite{Borexino_MSW}, $\sigma(E)$ the $^{136}$Xe$^{GS}$ $\rightarrow$ $^{136}$Cs$^{1^+}$ charged-current cross-section~\cite{Haselschwardt:2020ffr}, and $\epsilon (E)$ the $^{136}$Xe-CC signal efficiency (Figure~\ref{fig:ROC_effic}).

Using the ROC curves from the likelihood-based scintillation analysis developed in Sec.~\ref{sec:backgrounds}, we choose a point which achieves a background efficiency of $7.8\cdot 10^{-9}$ so that the total expected background rate in nEXO will be $\leq0.5$ events per $10$ years. The resulting signal efficiencies as a function of energy are shown in Figure~\ref{fig:ROC_effic} (right). These efficiencies are calculated as a function of the as-yet-unknown timing resolution, $\sigma$, of the scintillation readout, which degrades the signal efficiency for a fixed background rejection power. We calculate the expected number of $^{136}$Xe-CC events that nEXO will observe as a function of $\sigma$, shown in Figure~\ref{fig:nEXO_solar_nu_rates}. Choosing $\sigma=1$~ns as a reference point and assuming an HZ model, we find that nEXO can expect to observe on average $14.5$ CNO events, $134$ $^7$Be events, $12.5$ $pep$ events, and $1.7$ $^8$B events in $10$ years of runtime. 

\begin{figure}
    \centering
    \includegraphics[width=1\linewidth]{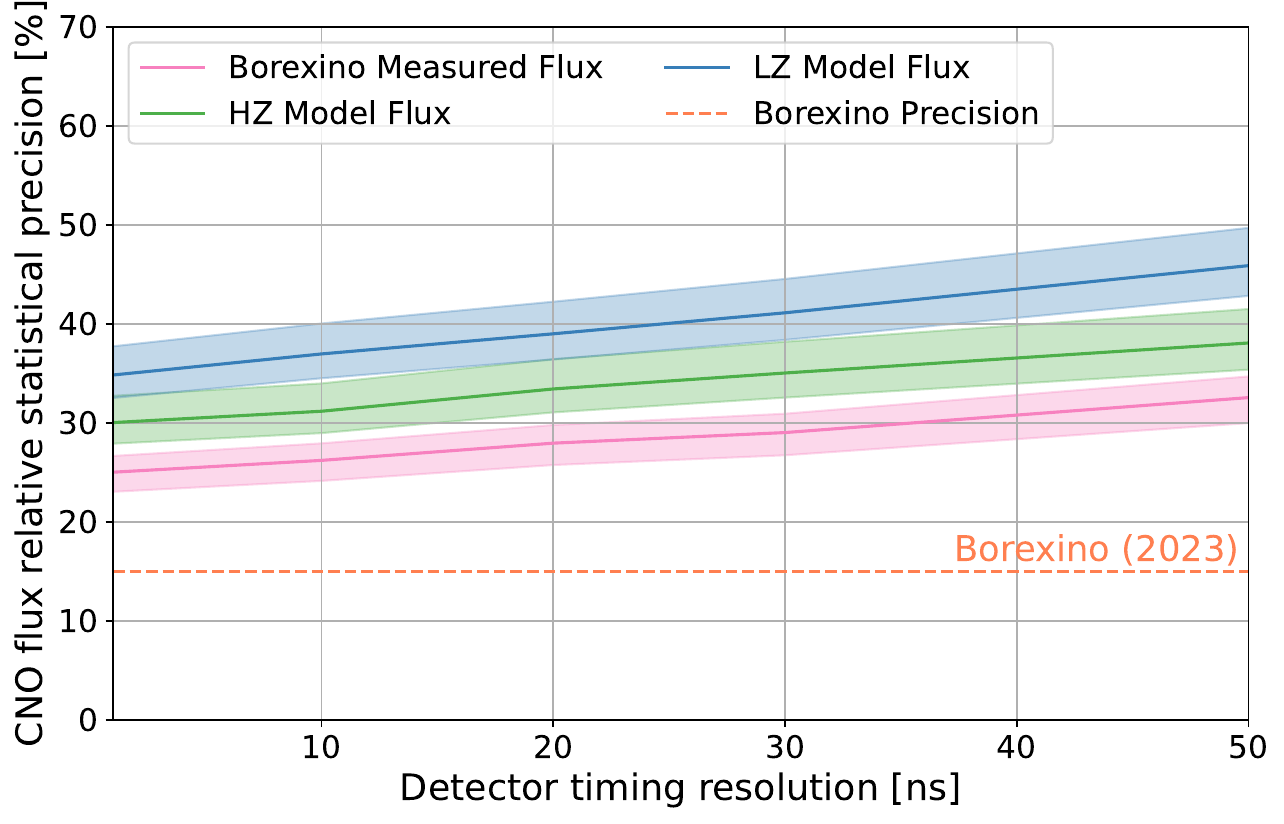}
    \caption{The statistical precision with which nEXO could measure the CNO neutrino flux after 10 years of runtime where separate curves are shown for the LZ, HZ, and Borexino measured fluxes. The colored bands around each curve indicate the 1$\sigma$ uncertainty. The Borexino precision is shown as a horizontal dashed line~\cite{Borexino,Gabriel}.}
    \label{fig:CNO-res}
\end{figure}

\begin{figure}
    \centering
    \includegraphics[width=1\linewidth]{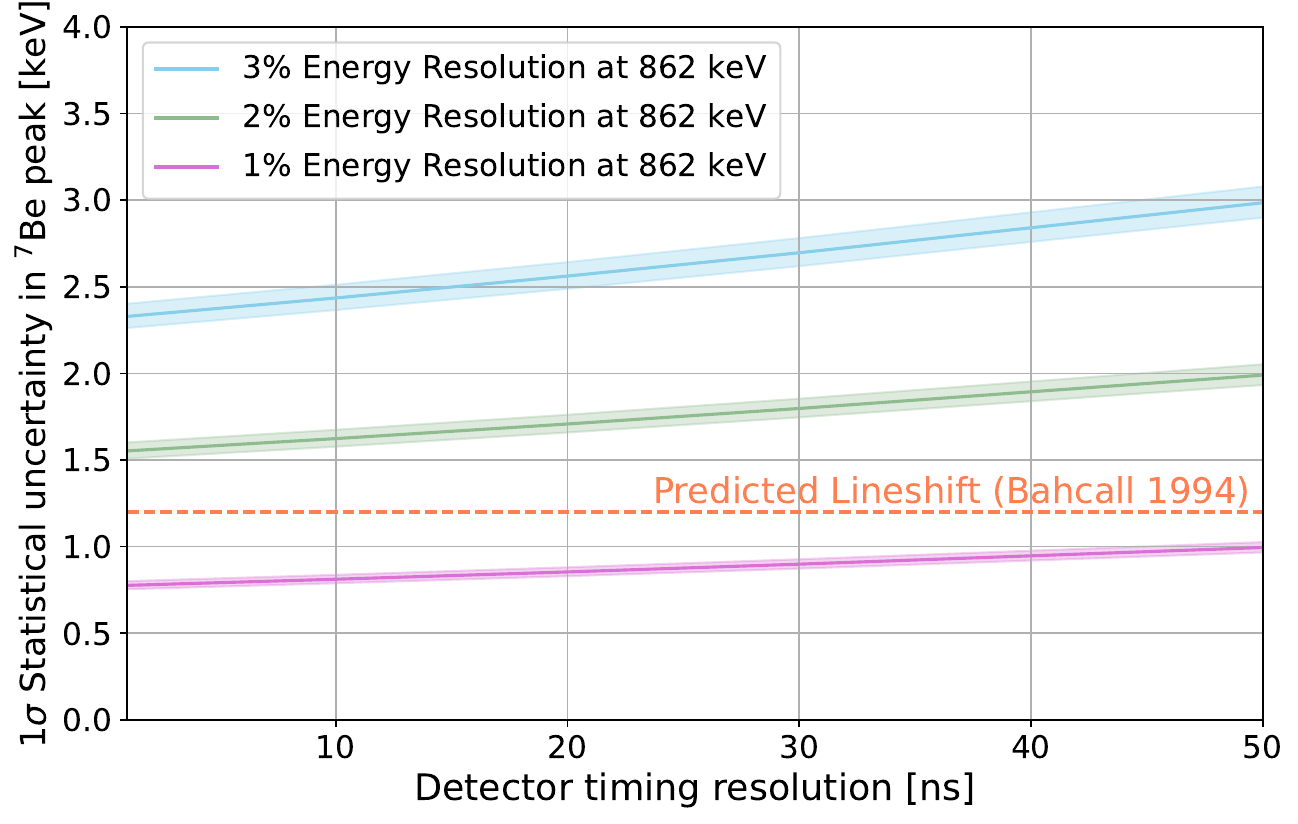}
    \caption{The expected precision with which nEXO could measure the $^7$Be line-shift assuming $\sigma=1$~ns timing resolution and the HZ solar flux. The predicted magnitude of this line-shift is $1.3$~keV and is denoted by the horizontal dashed line \cite{Bahcall:1994cf}.}
    \label{fig:Be-7-lineshift}
\end{figure}

\begin{figure}
    \centering
    \includegraphics[width=1\linewidth]{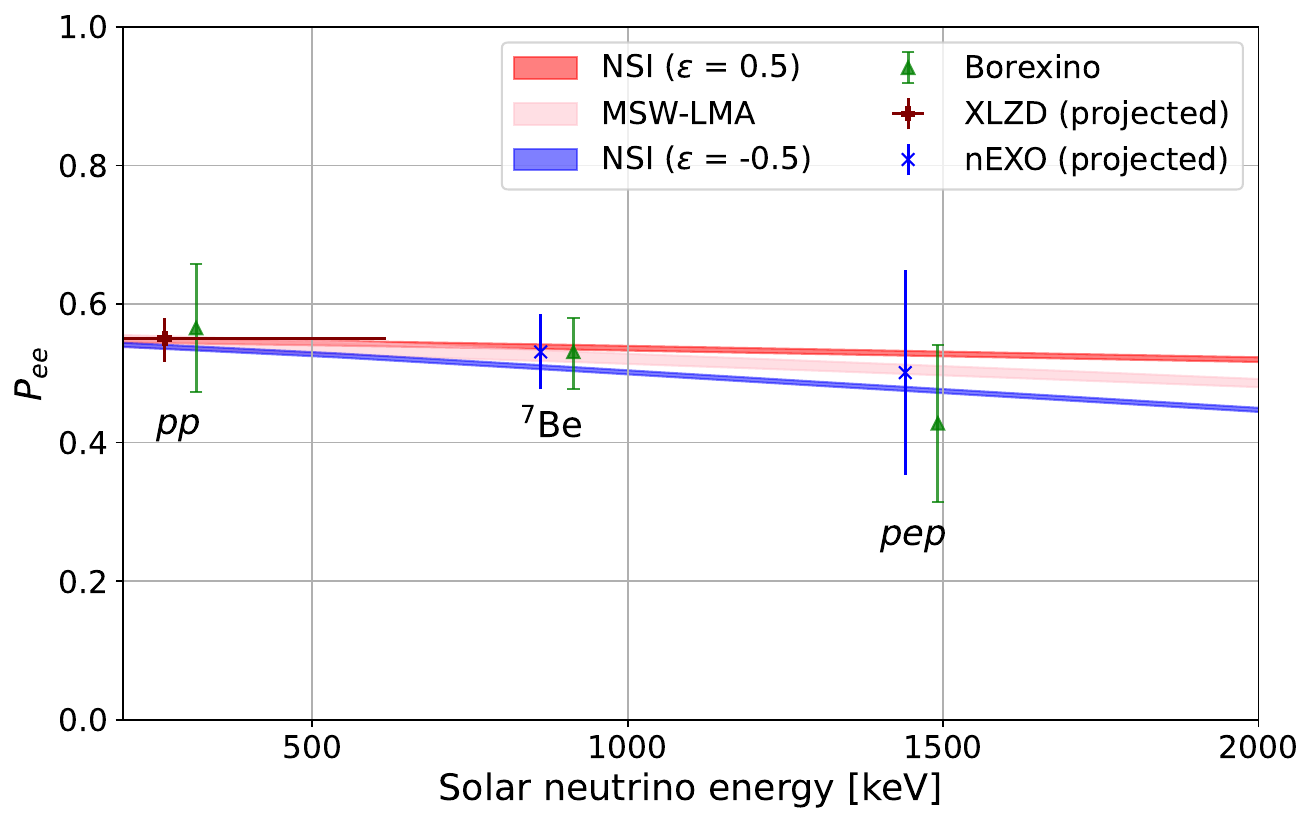}
    \caption{The expected precision with which nEXO can measure $P_{ee}$ for low energy solar neutrinos, along with the current measurement from Borexino~\cite{Borexino_MSW} (Borexino data points are artificially offset by 50 keV for clarity) and projected measurements from XLZD~\cite{XLZD}. nEXO projections assume $\sigma=1$~ns timing resolution and an HZ solar model~\cite{HZ_Model, Extra_HZ_Model}. These measurements constrain the presence of non-standard neutrino interactions (NSI). The pink curve indicates the MSW-LMA solution, while the red and blue curves indicate two examples of NSI~\cite{BOREXINO_pp, Borexino_MSW}.}
    \label{fig:P_ee}
\end{figure}

We then estimate the sensitivity of nEXO to various parameters of interest by generating $10,000$ toy MC spectra using our computed rates and the predicted spectra for each species of solar neutrino. Each toy is fit using an extended likelihood fit~\cite{ENLL} with free parameters for the number of events produced by each species of neutrino, as well as a parameter for the peak position (in energy) of the $^7$Be line. The results of this analysis, again reported as a function of $\sigma$, are illustrated in Figures~\ref{fig:CNO-res}, \ref{fig:Be-7-lineshift} \& \ref{fig:P_ee} and are discussed below.

Detecting CNO neutrinos is a significant experimental challenge due to their relatively low energy and flux~\cite{Theia, JUNO, Cerdeno:2017xxl}. To date, only the Borexino experiment has made a direct measurement of the CNO flux~\cite{borexinoCNO,Borexino}. Although the expected CNO event rate in nEXO  (shown in Figure~\ref{fig:nEXO_solar_nu_rates}) is significantly lower than that observed by Borexino---due to two orders of magnitude difference in target mass between the two detectors---the low background rate enables a measurement of the CNO flux which could potentially reach a statistical uncertainty of 25\%, only a factor of $\sim$2 less precise than Borexino's world-leading result. We illustrate these projections in Figure~\ref{fig:CNO-res}~\cite{Borexino}. Such a measurement would provide an independent confirmation of the Borexino results using an entirely new detection channel, and could improve our knowledge of the solar metallicity in a joint analysis.

nEXO will also be able to use $^{136}$Xe-CC interactions to measure $^7$Be solar neutrinos. While a measurement of solar metallicity in nEXO based solely on the $^7$Be flux is not expected to be competitive, such an analysis could supplement the CNO measurement. Furthermore, the charged-current interaction provides a powerful measurement of the average energy of the $^7$Be neutrinos; as mentioned previously, this provides a unique measurement of the temperature conditions in the solar core~\cite{Bahcall:1994cf}. The precision of this measurement depends not only on the timing resolution of the detector (which drives the detection efficiency), but also the detector's energy resolution at the $^7$Be energy. Most LXe TPCs can achieve $\mathcal{O}(1\%)$ resolution at these energy scales~\cite{XENON_e_res, LZ_E_res}; nEXO expects to achieve $\sim 2$\%, provided calibration related systematics can be sufficiently constrained~\cite{nEXO_sens}. We illustrate our projections for nEXO's precision in Figure~\ref{fig:Be-7-lineshift}. Assuming a $\sigma = 1$~ns timing resolution and the HZ solar flux, nEXO can make a measurement of the $^7$Be peak position with uncertainty $\pm 1.5$~keV, an order of magnitude more precise than the measurement from KamLAND ($\pm 16$~keV)~\cite{KamLAND_Be7}.

Finally, nEXO provides sensitivity to NSI through measurements of $P_{ee}(E)$~\cite{Borexino_MSW}. We illustrate nEXO's measurements with the $^{7}$Be and $pep$ components in Figure~\ref{fig:P_ee}. While nEXO will not be able to measure these effects at $pp$ energies due to the $^{136}$Xe-CC threshold (though the proposed XLZD LXe TPC could do so via electron-scattering (ES) channels~\cite{XLZD}), or $^8$B energies due to limited statistics,  the precision at $^7$Be and $pep$ energies offers sensitivity to NSI comparable to the current world-leading measurements from Borexino~\cite{Borexino_MSW}.

\section{\label{sec:DM} APPLICATION: FERMIONIC DARK MATTER}

We also study the prospects for nEXO to use $^{136}$Xe-CC interactions to search for fermionic dark matter (FDM) coupling to the SM. This interaction channel is sensitive to models in which the leptonic sector of the SM is embedded into a larger dark sector which contains a new dark gauge boson, $W'$, and an electron-flavor carrying dark matter particle, $\chi$, with an inert left-handed component and a right-handed component which completes the lepton right-handed doublets. Because of its electron-flavor, $\chi$ can couple to the SM through a new $\chi$-$W'$-$e^-$ vertex which gives rise to charged-current interactions analogous to Eq.~\ref{eq:CC} where $\nu_e$ is effectively replaced with $\chi$~\cite{Dror:2019onn, Second_Dror}. 

The charged-current interaction rate depends on the energy scale of the coupling, $\Lambda$, and the mass of the FDM particle, $m_{\chi}$. This rate in natural units ($\hbar=c=1$) is given as

\begin{equation}
    \label{eq:DM-Rate}
    R(m_{\chi}) =  N_{\mathrm{Xe}}n_{\mathrm{Xe}} \frac{ \rho_{\chi}} {2m_{\chi}} \mathcal{F}(Z+1, m_{\chi}-Q) \frac{|\vec{p}_e|}{16 \pi m_{\chi} M^2_{\mathrm{Xe}}}|M_{k}|^2
\end{equation}
where $N_{\mathrm{Xe}}$ is the number of $^{136}$Xe atoms in the active volume, $n_{\mathrm{Xe}}$ = $82$ the number of neutrons in the target isotope, $\rho_{\chi} = 0.3$~GeV/cm$^3$ the local dark matter density, $\mathcal{F}(Z+1,m_{\chi}-Q)$ the Fermi screening function, $\vec{p}_e$ the outgoing electron's $3$-momentum, $M_{\mathrm{Xe}}$ = $126.6$~GeV the mass of the $^{136}$Xe atom, and $M_{k}$ the parton spin averaged nucleon level matrix element~\cite{Dror:2019onn, Second_Dror}.

We compute nEXO's sensitivity to these signals by assuming no signal is present and calculating the upper limit on the observed event rate at a $95$\% confidence level (CL). We consider only FDM interactions which excite the $^{136}$Cs to its lowest lying $1^+$ state. If excitation to other states is appreciable, this will set a conservative limit as any higher energy states would add to the overall rate. Additionally, we confine our analysis to FDM candidates ranging in mass from \mbox{$m_{\chi}=E_{thresh}$}, to an upper-limit of \mbox{$m_{\chi}=S_n^{136Cs}+Q=6.9$}~MeV where \mbox{$S_n^{136Cs}=6.8$}~MeV is the neutron separation energy of $^{136}$Cs~\cite{CS_binding_energy, li2021exploration, zhang2021neutron}.

The expected FDM energy spectrum is modeled as a Gaussian centered at $E$ = $m_{\chi} - Q$ with a width given by the projected energy resolution of nEXO. This approximation is validated by simulating $10^6$ FDM signals in nEXO for various $m_{\chi}$ using the full Geant4-based \texttt{nexo-offline} framework~\cite{nEXO_sens, zepeng_sim}. 

Due to the powerful scintillation-based background discrimination, solar neutrino charged-current events are expected to be the dominant background for an FDM search with nEXO. Unlike backgrounds arising from ambient radioactivity, the solar neutrino background cannot be rejected on an event-by-event basis, since it will create the same event signature as an FDM interaction. Spectral information provides the only handle with which to discriminate against this background. 

\begin{figure}
    \centering
    \includegraphics[width=1\linewidth]{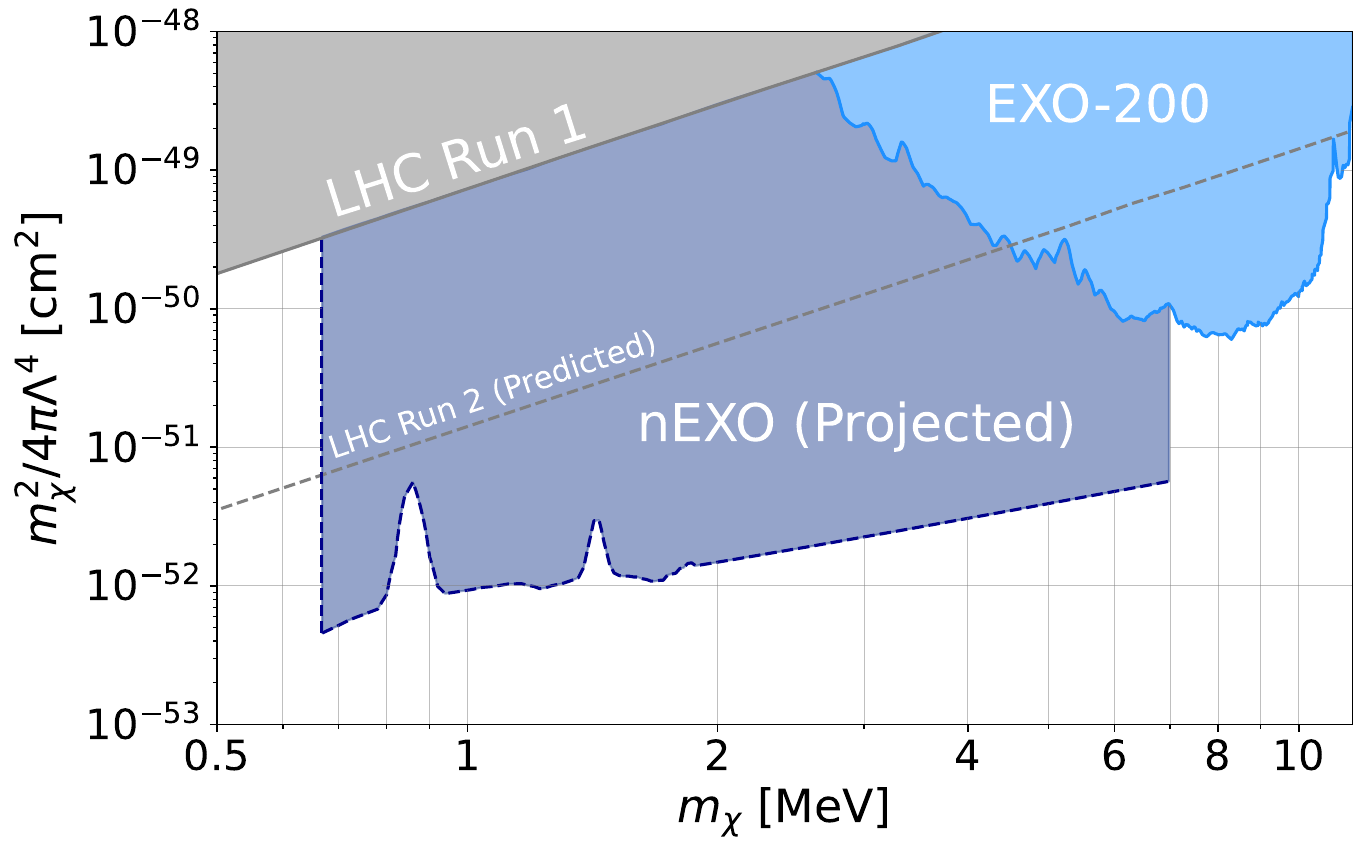}
    \caption{95\% CL sensitivity of nEXO to FDM charged-current interactions. LHC constraints are set by searches for $pp\rightarrow l\nu$, which constrain charged-current operators via constraints on helicity-non-conserving contact interaction models~\cite{CMS_8TeV}. Only the results for $1$~ns timing resolution are shown under the assumption of an HZ solar model background.}
    \label{fig:FDM-Exclusion}
\end{figure}

We simulate the expected solar neutrino background spectrum with no signal present, assuming an HZ solar model backround and $1$~ns timing resolution. An extended likelihood fit is applied to each simulated dataset~\cite{ENLL} and the $95\%$ confidence interval (CI) upper limit on the rate of dark matter detection is calculated using an ensemble of MC trials~\cite{Cowan}. The median upper limits are converted to cross-section upper limits using Eq.~\ref{eq:DM-Rate} and our results are shown in Figure~\ref{fig:FDM-Exclusion}.

Figure~\ref{fig:FDM-Exclusion} also shows the exclusion limits set by the EXO-200 LXe TPC~\cite{Ako} and the LHC at center of mass energies $\sqrt{s}=8$~TeV~\cite{Second_Dror, CMS_8TeV}. The EXO-200 analysis ($234.1$~kg$\cdot$yr exposure) performed a search similar to this work for charged-current interactions on Xe, considering all isotopes of Xe present in the detector ($\sim 80$\% $^{136}$Xe and $\sim 20$\% $^{134}$Xe) and all possible excited states of Cs, extending beyond the $m_{\chi} = S_n^{136}+Q$ limit of the present analysis. However, the analysis does not make use of the isomeric transitions, which were observed after its publication~\cite{Isomeric_Obs}, and thus is background-limited at lower masses. The LHC constraint is set by searches for missing energy in CMS proton-proton collisions, which produces constraints on the direct production of electron-coupling FDM. In addition to the region excluded by Run~1 of the LHC, the exclusion limit predicted to be set by Run~2~($\sqrt{s}=13$~TeV) is shown as a dashed line~\cite{Ma_LHC_Predicitions}. Although the Run~2 data has been published~\cite{CMS_13TeV, ATLAS_13TeV}, it has not yet been analyzed in the context of this model. In addition to the constraints shown in Figure~\ref{fig:FDM-Exclusion}, electron-coupling FDM can also be excluded based on cosmological constraints from FDM decay, though these limits become relevant at higher masses~\cite{FDM_Decay}.

The impact of the time-delayed coincident signal is most prominent at energies below $\sim5$~MeV, where naturally-occurring radioactive backgrounds are dominant. Due to the discrimination power of this signal channel, an electron-coupling FDM search in nEXO could expect more than three orders of magnitude improvement in sensitivity, and would extend the searchable parameter space by direct detection experiments to significantly lower masses. 
While these projections do depend on timing resolution, this effect is subdominant as they scale linearly with signal efficiency which changes by $\sim \times 2$ between $\sigma=1$ and $\sigma=50$~ns resolution (Figure~\ref{fig:ROC_effic} right). At higher energies, this analysis could be extended beyond $7$~MeV if the neutron separation channels of $^{136}$Cs are suppressed, as recent studies indicate~\cite{COHERENT:2023ffx}. Alternatively, the analysis could be modified to account for free neutrons in the detector as in Ref.~\cite{Sam} or could search for additional excited states of $^{136}$Cs and even $^{134}$Cs~\cite{Ako}. In these cases, one would forgo the background rejection of the time-delayed coincident signal and instead rely on the naturally low background rate at these energies, discussed in Sec.~\ref{sec:backgrounds}.

\begin{table*}[!t]
    \centering
    \small
    \renewcommand{\arraystretch}{1.3} 
    \setlength{\tabcolsep}{8pt} 
    \begin{tabular}{|c|c|c|c|}
        \hline
        \textbf{Measurement} & 
        \textbf{nEXO} & 
        \textbf{Origin-X}  & 
        \textbf{Current Result} \\ 
        \hline
        $\phi_{CNO}$  & $29\%$  &  $4\%$ & $15\%$ (Borexino \cite{Borexino})\\ 
        $\delta_{Be}$  & $\pm 1.5$~keV  & $\pm0.08$~keV & $\pm16$~keV (KamLAND \cite{KamLAND_Be7})\\ 
        $P_{ee}(E_{Be})$  & $10\%$  & $6\%$ & $9\%$ (Borexino \cite{BOREXINO_pp})\\ 
        $P_{ee}(E_{pep})$  & $29\%$ & $4\%$ & $26\%$ (Borexino \cite{BOREXINO_pp})\\
        $P_{ee}(E_B)$ & $78\%$ & $15\%$ &  $5\%$ (SNO \cite{SNO})\\
        \hline
    \end{tabular}
    \caption{The expected uncertainties for nEXO and an Origin-X style LXe TPC~\cite{ktonne} using $3$~ktons of natural LXe in a $7.4$~m cylindrical TPC compared to the current leading results assuming $\sigma=1$~ns timing resolution and HZ models~\cite{HZ_Model}. In the case of $P_{ee}(E_{Be})$, the Origin-X uncertainty is limited by the uncertainty in the overall flux of $^7$Be~\cite{Gabriel, BOREXINO_pp}.} 
    \label{tab:kton_stats}
\end{table*}

\section{\label{sec:ktonne} BEYOND THE TON SCALE}

In addition to nEXO, larger LXe TPCs are conceivable such as the upcoming generation of dark matter detectors~\cite{XLZD, PANDA-X} which are expected to have similar $^{136}$Xe exposure as nEXO. For this reason, such detectors might have comparable capabilities to nEXO in regards to the $^{136}$Xe-CC channel with the caveat that the larger size of these detectors will result in longer photon transport times~\cite{Darwin_Chroma}, which could impact their discrimination efficiency. Detailed optical transport simulations of these detectors will be needed to precisely estimate their capabilities. 

\begin{figure}
    \centering
    \includegraphics[width=\linewidth]{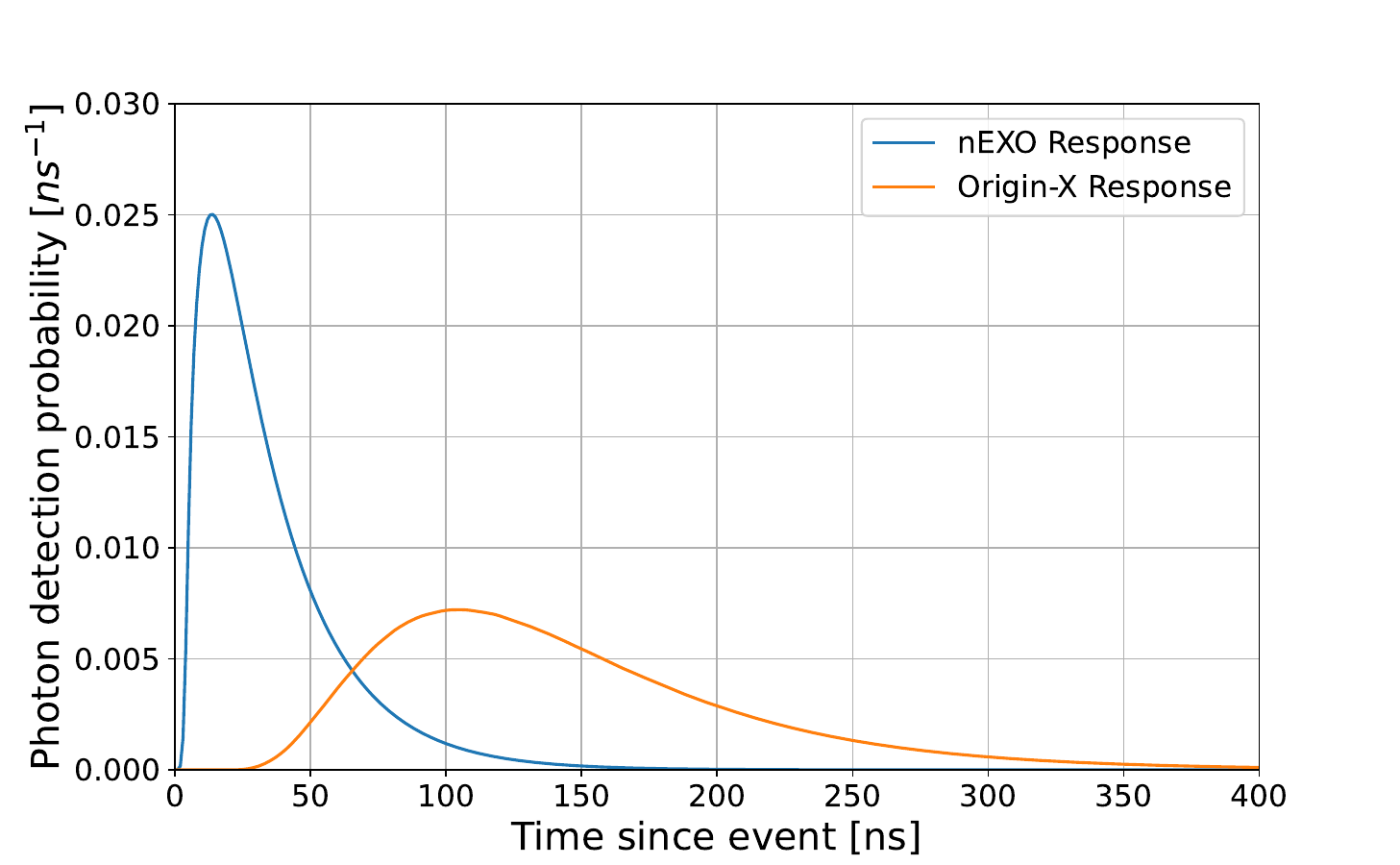}
    \caption{A comparison of the expected photon travel time in the nEXO detector and a $3$~kton natural xenon Origin-X detector with $\sigma=1$~ns timing resolution~\cite{ktonne, Ako_thesis}. The increased travel time results in $\sim35\%$ loss in signal efficiency compared to a nEXO scale LXe TPC.}
    \label{fig:kon-travel}
\end{figure}

Here, we consider beyond-the-next-generation experiments such as the Origin-X concept~\cite{llnlOriginx}, which would deploy either $3$~ktons of natural xenon, or $300$~tons of enriched xenon~\cite{ktonne, Anker}. Such a detector would have a dramatically increased exposure and would enable precision measurements of the solar neutrino spectrum via the $^{136}$Xe-CC channel. To account for the longer photon transit time, we use simulation results from Ref.~\cite{Ako_thesis}, which reports a Chroma simulation similar to that used in Sec.~\ref{sec:CC-Signal} but for a $7.4$~m LXe TPC. This simulation assumes an absorption length of $20$~m and a scattering length of $30$~cm for VUV photons~\cite{ktonne, Ako_thesis}. In such a large detector, Rayleigh scattering dominates the propagation of VUV photons, resulting in longer transit times and broader scintillation pulses than in the case of nEXO. We illustrate our simulated pulse shapes in Figure~\ref{fig:kon-travel}. We then incorporate the updated pulse shape for the Origin-X detector into our likelihood-based isomer tagging algorithm (Sec.~\ref{sec:CC-Signal}) and re-evaluate the signal/background discrimination.

The dominant background for such a detector is assumed to be $2\nu \beta \beta$ which will scale with $^{136}$Xe mass, while other backgrounds scale roughly with the surface area of the detector and will be strongly attenuated near the edges due to the self-shielding of the xenon. We therefore consider only $2\nu\beta\beta$ backgrounds and optimize the discrimination to provide background leakage to the level of $0.5$ events per $10$ years. Additionally, we assume that given the scale of future detectors, pile-up on scintillation channels (i.e. multiple photons interacting in the same sensor) will be negligible. Similarly, we omit the effects of (d)ExCT which has already been found to be close to negligible at the level of nEXO, and conservatively assume photosensors with the same PDE as the SiPMs for nEXO. 

We project that such a detector will achieve $\sim 35 \%$ worse signal efficiency than nEXO due to the extended scintillation pulse shape, but with $\sim \times 100$ the $^{136}$Xe exposure~\cite{ktonne}. The end result is that an Origin-X detector would observe $\sim 65$ times the $^{136}$Xe-CC event rate of nEXO at the same level of background leakage. 

We recalculate the expected $^{136}$Xe-CC event rates using the models in Sec.~\ref{sec:solar-nu} and show the results in Table.~\ref{tab:kton_stats}. Considering only statistical uncertainties, an Origin-X scale experiment could provide $>5 \sigma$ discrimination between the LZ and HZ solar models, definitively answering the solar metallicity question. Such a detector would also precisely measure the predicted $^7$Be line-shift and provide the most precise measurements of $P_{ee}$ at $^7$Be and $pep$ energies (as well as $pp$ energies via ES~\cite{XLZD}) to date. To make a CC measurement at $^8$B energies it will be necessary to extend the analysis beyond the neutron separation energy, as discussed in Sec.~\ref{sec:DM}.

\section{\label{sec:conclusion} CONCLUSION}

The observation of low-lying isomeric states in the nuclear structure of $^{136}$Cs has opened a new path for neutrino physics using detectors containing $^{136}$Xe. The time-delayed coincident signal created by the isomeric levels of $^{136}$Cs can enable background discrimination on the order of $10^{-9}$, which is more than sufficient to counterbalance the $2\nu \beta \beta$ backgrounds created by the $^{136}$Xe and, in fact, can allow LXe TPCs such as nEXO to perform searches with background rates below 1 event per 10 years.

In this work we demonstrate that nEXO is capable of studying solar neutrinos, in addition to its primary goal of $0 \nu \beta \beta$ detection, with precision approaching state-of-the-art. Specifically, the nEXO detector is predicted to measure the flux of CNO neutrinos and the survival probability of low energy solar neutrinos with statistical precision comparable to Borexino's world-leading results~\cite{borexinoCNO, BOREXINO_pp}. Additionally, because it will be able to reconstruct the energy of incoming solar neutrinos event-by-event, nEXO is expected to make an order of magnitude improvement on the measurement of the $^7$Be energy~\cite{KamLAND_Be7}, with $\sim1\sigma$ sensitivity to the predicted line-shift.

In addition to solar neutrino studies, the $^{136}$Xe-CC interaction channel is expected to allow nEXO to extend the searchable parameter space of LXe TPCs to electron-coupling FDM with mass as low as $668$~keV, and to search for almost three orders of magnitude smaller interaction strengths for these FDM models compared to current constraints~\cite{Ako, LHC, Second_Dror}. 

Beyond nEXO, $^{136}$Xe-CC studies in future generations of LXe TPCs hold great potential for pushing the boundaries of astroparticle physics. The next generation LXe dark matter detectors plan to have up to twice the $^{136}$Xe exposure as nEXO and so may be able to make measurements with precision comparable to or better than this work. A complete evaluation of these exciting possibilities will depend on the details of the detector's optical transport and cosmogenic backgrounds~\cite{XLZD, PANDA-X, Darwin_Chroma}. At the kiloton scale, a detector in the vein of the Origin-X concept~\cite{ktonne} could provide unprecedentedly precise measurements in multiple aspects of solar neutrino physics including a potential $5\sigma$ measurement of the $^7$Be line-shift~\cite{Bahcall:1994cf, Bahcall:1989ygn}, and---assuming uncertainty in the Gamow-Teller transition strength can be reduced \cite{GT, transitions}--a definitive answer to the solar metallicity problem via a precision measurement of the CNO neutrino flux~\cite{Vinyoles:2016djt}.

\section{ACKNOWLEDGMENTS}
G.R. was supported by a U.S. Department of Energy Science Graduate Student Research fellowship as part of contract number DE-SC0014664. This work was supported in part by Laboratory Directed Research and Development (LDRD) programs at Brookhaven National Laboratory (BNL), Lawrence Livermore National Laboratory (LLNL), Pacific Northwest National Laboratory (PNNL), and SLAC National Accelerator Laboratory. The authors gratefully acknowledge support for nEXO from the Office of Nuclear Physics within DOE’s Office of Science under grants/contracts DE-AC02-76SF00515, DE-AC05-76RL01830, DE-AC52-07NA27344, DE-FG02-01ER41166,  DE-FG02-93ER40789, DE-SC0012654, DE-SC0012704, DE-SC0014517, DE‐SC0017970, DE-SC0020509, DE-SC0021383, DE-SC0021388, DE-SC002305, DE-SC0024666, DE-SC0024677 and support by the US National Science Foundation grants NSF PHY-2011948 and NSF PHY-2111213; from NSERC SAPPJ-2022-00021, CFI 39881, FRQNT 2019-NC-255821, and the CFREF Arthur B. McDonald Canadian Astroparticle Physics Research Institute in Canada; from IBS-R016-D1 in South Korea; and from CAS in China.

\FloatBarrier
\bibliographystyle{apsrev4-2}
\bibliography{main}

\end{document}